\documentclass[conference]{IEEEtran}
\IEEEoverridecommandlockouts{}
\usepackage[top=50pt, left=54pt, right=54pt, bottom=58pt]{geometry}
\usepackage{cite}
\usepackage{amsmath,amssymb,amsfonts}
\usepackage{algorithmic}
\usepackage{graphicx}
\usepackage{textcomp}
\usepackage{xcolor}
\usepackage{hyperref}
\usepackage{diffcoeff}
\usepackage{booktabs}
\usepackage{siunitx}

\def\BibTeX{{\rm B\kern-.05em{\sc i\kern-.025em b}\kern-.08em
 T\kern-.1667em\lower.7ex\hbox{E}\kern-.125emX}}
\newcommand{\ui}[2]{#1_{\mathrm{#2}}}

\newcommand{\coo}{\ensuremath{\mathrm{CO_2}}}
\newcommand{\chho}{\ensuremath{\mathrm{CH_2O}}}
\newcommand{\citep}{\cite}

\begin{document}

\title{\vspace*{18pt}Carbon Neutral Greenhouse: Economic Model Predictive Control Framework for Education
    \thanks{The authors acknowledge the contribution of the Program to support young researchers at STU under the project Adaptive and Robust Change Detection for Enhanced Reliability in Intelligent Systems. The authors gratefully acknowledge the contribution of the Scientific Grant Agency of the Slovak Republic under the grants 1/0490/23 and 1/0297/22. This research is funded by the Horizon Europe under the grant no. 101079342 (Fostering Opportunities Towards Slovak Excellence in Advanced Control for Smart Industries).}
}

\author{\IEEEauthorblockN{Marek Wadinger, Rastislav F\'aber, Erika Pavlovi\v cov\'a, Radoslav Paulen}
    \IEEEauthorblockA{\textit{Faculty of Chemical and Food Technology,}
        \textit{Slovak University of Technology in Bratislava,}
        \textit{Bratislava, Slovakia} \\
        \texttt{marek.wadinger@stuba.sk}}
}

\maketitle

\begin{abstract}
    This paper presents a comprehensive framework aimed at enhancing education in modeling, optimal control, and nonlinear Model Predictive Control~(MPC) through a practical greenhouse climate control model. The framework includes a detailed mathematical model of lettuce growth and greenhouse, which are influenced by real-time external weather conditions obtained via an application programming interface~(API). Using this data, the MPC-based approach dynamically adjusts greenhouse conditions, optimizing plant growth and energy consumption and minimizing the social cost of CO\textsubscript{2}. The presented results demonstrate the effectiveness of this approach in balancing energy use with crop yield and reducing CO\textsubscript{2} emissions, contributing to economic efficiency and environmental sustainability.
    Besides optimizing lettuce production, the framework also provides a valuable resource for making control systems education more engaging and effective. The main aim is to provide students with a hands-on platform to understand the principles of modeling, the complexity of MPC and the trade-offs between profitability and sustainability in agricultural systems. This framework provides students with hands-on experience, helping them to understand the control theory better, connecting it to the practical implementation, and developing their problem-solving skills. The framework can be accessed at \url{ecompc4greenhouse.streamlit.app}.
\end{abstract}


\section{Introduction}

In the ever evolving field of engineering, there is an increasing demand for graduates possessing not only theoretical knowledge but also practical skills applicable in real-world scenarios. Technical university education requires a solid foundation in theory coupled with opportunities for practical learning, often gained through participation in projects, solving real-world problems, and conducting experiments. The benefits of interactive tools in control theory instruction have been highlighted in previous works~\cite{Emami1991, Guzman2013}. Practical learning does not always rely on physical experiments with costly lab equipment. Simulations and interactive tasks can serve as effective alternatives, offering new learning opportunities.

The farming today is labor-intensive, seasonal, constrained by irrigation, and rely on subsidized inputs, leading to environmental issues such as eutrophication, deforestation and soil degradation. With nearly 70\% of global water resources consumed by agriculture~\cite{Debroy2024}, greenhouses offer a solution by providing controlled environments that enhance productivity beyond what open-field cultivation can achieve. However, these systems face challenges, such as fluctuating internal temperatures that can harm crops. Effective climate management, through controlled ventilation and heating, is essential for maintaining optimal conditions and improving yields~\cite{Wu2019}.

Solar radiation is vital for plant growth and energy accumulation in greenhouses. Key metrics include Global Horizontal Irradiance~(\(GHI\)) and Photosynthetically Active Radiation~(\(PAR\)), the latter directly influencing photosynthesis. A recent review highlighted various empirical models estimating PAR in areas with scarce measurements, emphasizing the relationship between PAR and GHI~\citep{NoriegaGardea2021}. Prediction accuracy of models prediicting has been enhanced, while also addressing the lack of historic performance data is lacking~\cite{Iddio2020, MaLu2022}. Other contributions apply deep learning to predict plant growth using time-series data in forecasting future growth~\cite{rs13030331}.

Various control methods have also been explored to enhance greenhouse environments and resource use efficiency. Adaptive control adjusts parameters based on real-time feedback, enabling dynamic responses to changing conditions~\cite{Tian2022}. Nonlinear feedback control addresses system complexity through advanced algorithms, improving overall system responsiveness and stability~\cite{Bood2023}. Fuzzy control effectively manages imprecise data and uncertainties, making it suitable for fluctuating environmental variables~\cite{smartcities7030055}. Robust control ensures stability despite external disturbances, contributing to reliable operation under varying conditions~\cite{Zhang2021}. Optimal control fine-tunes actions to achieve the best outcomes under specific constraints~\cite{Debroy2024, SVENSEN2024108578}. Despite their strengths, these approaches can lead to complex implementation with frequent adjustments, leading to higher energy consumption and wear on actuators.

While PID controllers are valued for their simplicity and effectiveness, IoT and machine learning are also being integrated into greenhouse control, as demonstrated by~\cite{Wang2024}, who combined these technologies with PID for real-time monitoring. Nonetheless, managing greenhouse systems using PID can be challenging due to the need for multiple controllers and extensive tuning. This process is time-consuming and case-dependent, often lacking guaranteed optimal results. Consequently, MPC has emerged as a preferred approach for greenhouse climate control~\cite{Hu2022}, enabling continuous adjustment of setpoints through sample-by-sample online optimization.

Integrating advanced control techniques into education has become increasingly important. Recent studies~\cite{WangEducation2024, Zakova2024} have shown that web-based simulation platforms significantly enhance students' problem-solving skills by bridging the gap between theory and practice. The presented framework aims to provide a similar educational experience within greenhouse climate control, focusing on modeling, optimal control, and Nonlinear Economic Model Predictive Control~(NEMPC).

This study presents a web interface to optimal greenhouse climate control, to achieve enhanced crop yields, energy efficiency and reduction in \coo\ emissions. Utilizing principles of thermodynamics, fluid dynamics, and mass transfer, along with weather and carbon intensity forecasts, we developed a simulation environment employed by NEMPC to adjust ventilation, heating, humidification, and \coo\ enrichment in real-time. Beyond technical contributions, the framework serves as an educational tool, allowing students to engage with explore the economic and sustainability trade-offs in agricultural systems.

\section{Greenhouse Climate Model}\label{sec:greenhouse}
In this section, we provide a mathematical model of the greenhouse environment, focusing on wind, temperature, humidity, and \coo\ concentration dynamics implemented within the GES software~\cite{rmward61_2019} based on the Gembloux Dynamic Greenhouse Climate Model~\cite{GDGCM} and the work by Vanthoor~\cite{Vanthoor2011}. For the complete list of parameters and their values with references, see the source code repository\footnote{\url{https://github.com/MarekWadinger/ecompc-greenhouse-platform/blob/main/core/greenhouse_model.py}}.

\begin{figure}
    \centering
    \includegraphics[width=\linewidth]{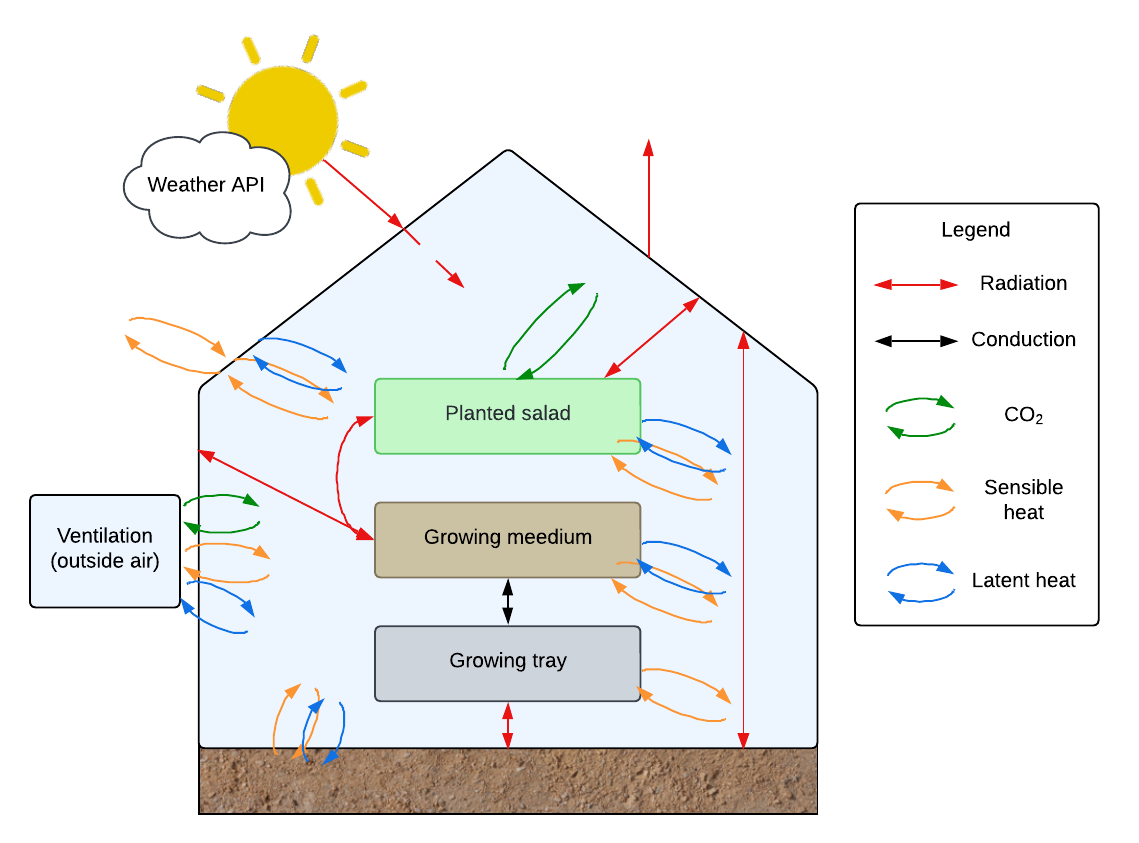}
    \caption{A simplified diagram of the heat, mass and \coo\ exchanges modelled within the framework. Image adapted from~\cite{rmward61_2019}.}\label{fig:diagram}
\end{figure}

\subsection{Temperature Dynamics}\label{subsec:temperature}

The temperature dynamics inside the greenhouse is modeled by considering the energy exchange due to convection, radiation, and conduction --- Fig.~\ref{fig:diagram}. The temperatures of different compartments (cover, internal air, vegetation, tray, etc.) are described using the following equations.

The convective heat transfer between two surfaces reads as:
\begin{equation}
    \text{Nu} = \max \left( \text{Nu}_G, \text{Nu}_R \right),
\end{equation}
where \(\text{Nu}_G\) and \(\text{Nu}_R\) are the Nusselt numbers for free and forced convection, respectively:
\begin{align}
    \text{Nu}_G & = 0.5  {\left( \frac{\text{Gr}}{10^5} \right)}^{0.25} + 0.13  {\left(\frac{\text{Gr}}{10^5}\right)}^{0.33}, \\
    \text{Nu}_R & = 0.6  {\left(\frac{\text{Re}}{20000}\right)}^{0.5} + 0.032  {\left(\frac{\text{Re}}{20000}\right)}^{0.8},
\end{align}
where Re represents the Reynolds number, and Gr represents the Grashof number. The convective heat flux is then:
\begin{equation}
    Q_{\text{conv}} = A_{\text{c}}\, \text{Nu}\, \lambda_{\text{air}} \frac{T_1 - T_2}{d_{\text{c}}},
\end{equation}
where \(A_c\) and \(d_c\) represent the characteristic area and length of a compartment, respectively, and \(\lambda_{air}\) is the approximate air thermal conductivity at room temperature.

The radiative heat transfer is described by:
\begin{equation}
    Q_{\text{rad}} = \frac{\varepsilon_1  \varepsilon_2}{1 - \rho_1  \rho_2  F_{12}  F_{21}}  \sigma  A_1  F_{12}  \left( T_1^4 - T_2^4 \right),
\end{equation}
where \(\sigma \) is the Stefan-Boltzmann constant, \(F_{12}\) is the view factor from surface 1 to surface 2, \(T_1\) and \(T_2\) represent the temperatures of the surfaces, and \(\varepsilon \) and \(\rho \) are the emissivity and reflectivity of the surfaces, respectively.

The conductive heat transfer through a medium is given by:
\begin{equation}
    Q_{\text{cond}} = \frac{A \lambda_{\text{c}}}{d_\text{l}} (T_1 - T_2),
\end{equation}
where \(\lambda_{\text{c}} \) is the thermal conductivity of a compartment, and \(d_{\text{l}}\) is the thickness of the conducting layer.

\subsection{Humidity Dynamics}\label{subsec:humidity}

The greenhouse humidity is modeled by considering the mass transfer of vapor. The air moisture content is given by:
\begin{equation}
    C_w = \rho_{\text{air}} \exp^{11.56 - \frac{4030}{T + 235}},
\end{equation}
where \(\rho_{\text{air}}\) is the density of air.


\subsection{Carbon Dioxide Concentration Dynamics}

The \coo\ concentration in the greenhouse is affected by photosynthesis and external conditions. The external \coo\ concentration is computed as:
\begin{equation}
    C_{\text{ext}} = \frac{4 \times 10^{-4}  M_c  P_{\text{atm}}}{R  T_{\text{ext}}},
\end{equation}
where \(M_c\) is the molar mass of \coo, \(P_{\text{atm}}\) is the atmospheric pressure, \(R\) is the gas constant, and \(T_{\text{ext}}\) is the external air temperature in Kelvin.

The internal \coo\ concentration in ppm is given by:
\begin{equation}
    C_{\text{int, ppm}} = \frac{C_c  R  T_i}{M_c  P_{\text{atm}}} \times 10^6,
\end{equation}
where \(C_c\) is the \coo\ density, and \(T_{\text{in}}\) is the internal air temperature in Kelvin.

The greenhouse climate model integrates the models of temperature, humidity, and \coo\ concentration into a dynamic system represented by a state vector \(x\) = [\(T_c, T_i, T_v, T_m, T_p, T_f, T_s, C_w, C_c, x_{\text{sdw}}, x_{\text{nsdw}}\)] \( \in\mathbb R^{11} \), and input vector \(u\) = [\(Q_{\text{heater}}, R_{\text{fan}}, V_{\text{humid}}, G_{\coo} \)] \( \in\mathbb R^{4} \) for the heating power, ventilation, humidification, and \coo\ enrichment. The \(T_c\) represents the cover temperature, \(T_i\) the internal air temperature, \(T_v\) the plant temperature, \(T_m\) the growing medium temperature, \(T_p\) the tray temperature, \(T_f\) the floor temperature, and \(T_s\) the temperature of the soil layer. Additionally, \(C_w\) denotes the density of water vapor, \(C_c\) the \coo\ density, \(x_{\text{sdw}}\) the structural dry weight of the plant, and \(x_{\text{nsdw}}\) the non-structural dry weight of the plant.

The model is influenced by external conditions \( p = \left[
T_{\text{ext}}, T_{\text{app}}, v_{\text{wind}}, H_{\text{rel}}, \mathbf{I}_{\text{POA,direct}}\in \mathbb{R}^{8}, \mathbf{I}_{\text{POA,diffuse}}\in \mathbb{R}^{8}
\right] \in \mathbb{R}^{20} \). The \(T_{\text{ext}}\) represents the external air temperature, \(T_{\text{app}}\) the apparent temperature, \(v_{\text{wind}}\) the wind speed, \(H_{\text{rel}}\) the relative humidity, and \(\mathbf{I}_{\text{POA,direct}}\) and \(\mathbf{I}_{\text{POA,diffuse}}\) the POA (plane of array) direct and diffuse solar radiation, respectively. Additionally, \(\mathbf{I}_{\text{POA,direct}}\) and \(\mathbf{I}_{\text{POA,diffuse}}\) from all planes of the greenhouse are integrated, providing a realistic and dynamic input for the control strategy.

\subsection{Actuation Control Systems}
Actuators operate by converting an input control signal from the system into the appropriate mechanical action to regulate environmental variables~\cite{Butterfield2018}. The actuators are used to control temperature, humidity, ventilation, and \coo\ concentration, summarized in Table~\ref{tab:actuators}. We monitor each actuator's contribution to the overall energy balance, operating costs, and \coo\ emissions.

\begin{table}
    \centering
    \caption{Summary of Actuator Model Parameters}\label{tab:actuators}
    \begin{tabular}{lcc}
        \toprule
        Actuator        & Label                   & Unit                   \\
        \midrule
        Heater          & \( Q_{\text{heater}} \) & W                      \\
        Fan             & \( R_{\text{fan}} \)    & m\textsuperscript{3}/s \\
        Humidifier      & \( V_{\text{humid}} \)  & l/h                    \\
        \coo\ Generator & \( G_{\coo} \)          & kg/h                   \\
        \bottomrule
    \end{tabular}
\end{table}

We model the actuators by adjusting the control signal, \( u(t) \), which ranges from 0 to 100\%, where 0\% and 100\% corresponds to no actuation and maximum actuation, respectively. The actuation level, \( a(u) \), is then calculated as:
\begin{equation}
    a(u) = \frac{u}{100}  a_{\text{max}},
\end{equation}
where \( a_{\text{max}} \) represents the maximum of the specific actuator.

The power consumption of an actuator is determined by:
\begin{equation}
    P(u) = \frac{p_{\text{unit}}}{\eta}  a(u),
\end{equation}
where \( p_{\text{unit}} \) is the power per unit of actuation, and \( \eta \) represents the efficiency of the actuator.

The total energy cost in EUR is calculated as:
\begin{equation}
    C_{\text{energy}}(u) = \frac{E_{\text{cost}}  \Delta t}{1000 \times 3600}  P(u),
\end{equation}
where \( E_{\text{cost}} \) is the cost of energy in EUR per kWh, and \( \Delta t \) is the time step in seconds.

The \coo\ emissions generated by an actuator are given by:
\begin{equation}
    E_{\coo}(u) = \frac{I_{\coo}  \Delta t}{1000 \times 3600}  P(u),
\end{equation}
where \( I_{\coo} \) is the carbon intensity in g\coo\ eq/kWh~\cite{ElectricityMaps2022}. The associated cost of these emissions is:
\begin{equation}
    C_{\coo}(u) = C_{\coo\text{cost}}  E_{\coo}(u),
\end{equation}
where \( C_{\coo\text{cost}} \) is the social cost of \coo\ in EUR/g\coo\ eq.

The heating power is determined based on the temperature setpoint, \( T_{\text{sp}} \), and air volume of the greenhouse, \( \Omega \), as:
\begin{equation}
    Q_{\text{heater}} = \rho_{\text{air}}  c_{\text{air}}  \Omega  (T_{\text{sp}} - T_{\text{ambient}})  \frac{Q_{\text{air}}}{3600},
\end{equation}
where \( \rho_{\text{air}} \) is the air density, \( c_{\text{air}} \) is the air specific heat capacity, and \( Q_{\text{air}} \) represents the fresh air exchange rate.

The ventilation rate, \( R_{\text{fan}} \), is based on the air changes per hour (\( \text{ACPH} \)) --- the number of times the air in a space is completely replaced in an hour --- and greenhouse volume \( \Omega \):
\begin{equation}
    R_{\text{fan}} = \Omega \frac{\text{\( \text{ACPH} \)}}{3600}.
\end{equation}

The humidification rate, \( V_{\text{humid}} \), is calculated as:
\begin{equation}
    V_{\text{humid}} = \Omega \frac{\phi_{a, 80 - 40}}{\rho_{\text{water}}},
\end{equation}
where and \( \rho_{\text{water}} \) is the water density and  \( \phi_{a, 80 - 40} \) is the maximum change in absolute humidity per hour which corresponds to difference between the absolute humidity at 80\% and 40\% relative humidity at a temperature of 20\( ^\circ \)C.

The \coo\ generation is based on the desired change rate of the \coo\ density, \( \dot{c}_{co2} \) as:
\begin{equation}
    G_{\coo} = \dot{c}_{co2}  V.
\end{equation}

\section{Plant Growth Model}\label{sec:lettuce_growth}
The growth of plant is modeled using a dynamic system of equations that captures the behavior of both structural (SDW) and non-structural dry weight (NSDW) of the plant~\cite{VANHENTEN199455}. The dynamics of conversion of  model is influenced by environmental factors such as temperature, light (PAR), and \coo\ concentration. The model equations are parameterized using well-established constants from the literature and adapted for dynamic simulation. For the complete list of parameters and their values with references, see source code repository\footnote{\url{https://github.com/MarekWadinger/ecompc-greenhouse-platform/blob/main/core/lettuce_model.py}}.

\subsection{Growth Dynamics} The rate of change of structural dry weight is governed by the specific growth rate \( \ui{r}{gr} \) and is modeled as:

\begin{equation}
    \diff{\ui{x}{sdw}}{t} = \ui{r}{gr}(T_i) \ui{x}{sdw}
\end{equation}
\begin{equation}
    \begin{aligned}
        \diff{\ui{x}{nsdw}}{t} & = c_{\chho/\coo} \ui{f}{phot}(C_c, T_i) - \ui{r}{gr}(T_i) \ui{x}{sdw} - \ui{f}{resp}(T_i)    \\
                               & \quad - \left( \frac{1 - Y_{\chho/\coo}}{Y_{\chho/\coo}} \right) \ui{r}{gr}(T_i) \ui{x}{sdw}
    \end{aligned}
\end{equation}

where \( c_{\chho/\coo} \) is the conversion factor representing the efficiency of absorbed carbon dioxide conversion into biomass, \( Y_{\chho/\coo} \) is the yield, \( \ui{f}{phot} \) is the gross photosynthesis rate, and \( \ui{f}{resp} \) is the maintenance respiration rate.

\subsection{Specific Growth Rate} The specific growth rate \( \ui{r}{gr} \) describes the accumulation rate of the structural biomass (SDW) in response to the available non-structural dry weight (NSDW). This process is temperature-dependent and scales based on plant temperature sensitivity. The NSDW provides a reservoir of energy for structural growth. The efficiency of the energy conversion is central to the plant's overall biomass accumulation.

\subsection{Maintenance Respiration} The maintenance respiration rate \( \ui{f}{resp} \) accounts for the energy expenditure required to sustain the plant's basic metabolic functions. The respiration process is divided between the shoot and root components, each part having its own maintenance respiration rate, scaled to the plant dry mass. Respiration demands increase as temperature rises. It is a key factor in determining the amount of energy for growth as it consumes a portion of the energy generated from photosynthesis.

\subsection{Gross Photosynthesis} The gross canopy photosynthesis rate \( \ui{f}{phot} \) represents the total \coo\ assimilation by the plant. This is primarily driven by the maximum \coo\ assimilation rate, which depends on the incident PAR, \coo\ concentration, and the canopy's light-use efficiency. The leaf area index and extinction coefficient further influence the efficiency of light absorbance across the canopy. Canopy conductance plays a critical role in this process by regulating the rate at which \coo\ diffuses into the leaf. The combined effect of boundary layer conductance, stomatal conductance, and carboxylation conductance ensures that \coo\ reaches the sites of photosynthesis efficiently. Carboxylation conductance itself is temperature-dependent, reflecting the enzymatic activity being affected by ambient conditions.

\section{Nonlinear Economic Model Predictive Control}\label{sec:mpc}
The NEMPC is adopted to maximize the profit of growing lettuce in a greenhouse. This is achieved by maximizing lettuce yield while minimizing operating costs over a finite prediction horizon. The NEMPC framework incorporates the dynamics of the greenhouse and time-varying external conditions.

\subsection{System Dynamics}\label{subsec:mpc_dynamics}

The greenhouse system dynamics is modeled by a set of nonlinear equations (see Section~\ref{sec:greenhouse}) as:
\begin{equation}
    x(t+1) = f\left( x(t), u(t), p(t) \right),
\end{equation}
where, at time step \(t\), \(x(t) \in \mathbb{R}^{n_x}\) represents the state vector, \(u(t) \in \mathbb{R}^{n_u}\) is the control input vector (actuators), and \(p(t)\) are time-varying parameters, i.e., external climate conditions.

\subsection{Economic Objective Function}\label{subsec:mpc_objective}

The goal of NEMPC is to maximize the revenue from lettuce production while minimizing the costs associated with actuator use. The objective function is composed of two terms: the profit from biomass accumulation and the actuating costs.

The revenue from lettuce production is proportional to the change in biomass between the initial state~\(x(0)\) and the current state~\(x(t)\), expressed as:
\begin{equation}
    R(t) = \frac{P_L A_c}{\rho_{dw}} \textstyle\sum_{i\in \{ \text{sdw}, \text{nsdw} \}}(x_i(t) - x_i(0)),
\end{equation}
where \(P_L\) is the price of lettuce per gram, \(A_c\) is the cultivated area, and \(\rho_{dw}\) stands for dry-to-wet ratio.

The actuating cost at each time step is given by:
\begin{equation}
    C_u(t) = \textstyle\sum_{i} C_{\text{energy}}(u_i(t)) + C_{\coo}(u_i(t)).
\end{equation}

The total cost at each time step is:
\begin{equation}
    l_t = -R(t) + C_u(t).
\end{equation}

\subsection{Optimization Problem Formulation}

The objective of the NEMPC is to minimize the cumulative stage cost over a finite prediction horizon \(N\), subject to system dynamics and constraints. The problem is formulated as:
\begin{equation}
    \min_{{\{u(t)\}}_{t=0}^{N-1}} \sum_{t=0}^{N-1} l_t(x(t), u(t)),
\end{equation}
subject to:
\begin{align}
    \text{s.t. } & x(t+1) = f(x(t), u(t), p(t)),                                      \\
                 & u_{\min} \leq u(t) \leq u_{\max}, \quad \forall t = 0, \dots, N-1, \\
                 & x_{\min} \leq x(t) \leq x_{\max}, \quad \forall t = 0, \dots, N,   \\
                 & x(0) = x_{\text{initial}}.
\end{align}
Here, \(x_{\min}\) and \(x_{\max}\), respectively, \(u_{\min}\) and \(u_{\max}\) represent the bounds on the state, respectively, control input variables.

\subsection{Time-Varying Parameters}

The external climate conditions are modeled as time-varying parameters \( p(t) \). These include outdoor temperature, solar radiation, and humidity, all provided by real-time weather data.

\section{Educational Web Interface}
\begin{figure}\label{fig:web}
    \centering
    \includegraphics[width=\linewidth, trim=50 50 50 50]{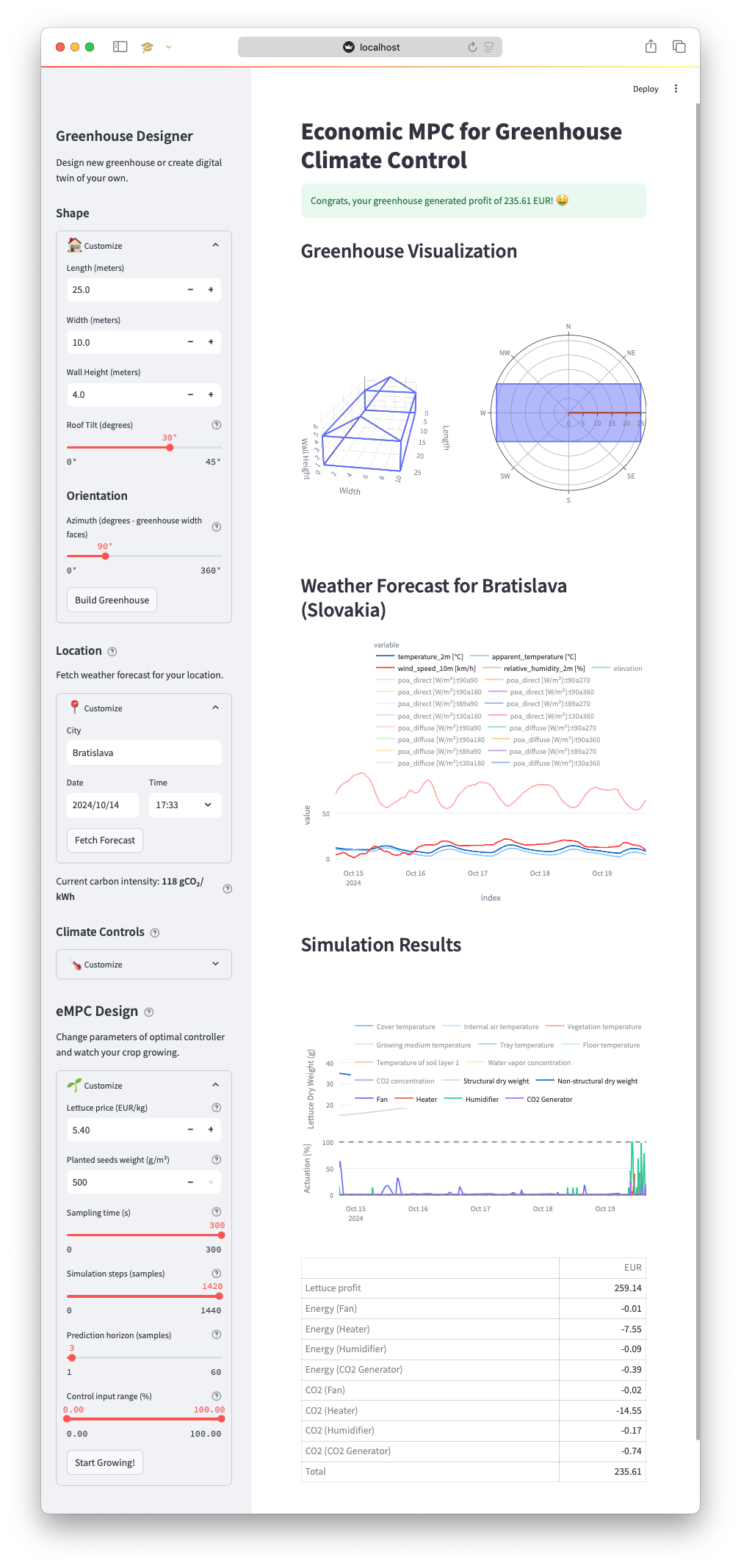}
    \caption{Education web interface. The sidebar ---  customization layers; the main window --- greenhouse structure, orientation, weather forecast, simulation results, and cost-profit analysis.}
\end{figure}

Figure~\ref{fig:web} demonstrates the interactive web-based educational tool. Via four layers of customization of the greenhouse, users may learn the main benefits of optimal control and challenges related to non-linearity. The first layer is the greenhouse structure, where the user can select the shape of the greenhouse, affecting the energy exchange with the environment and suggested scaling of the actuation units.

The second layer sets the orientation and location of the greenhouse. Here, we establish a connection to weather and carbon intensity forecast APIs to provide real-time data along with forecasts and history replays.
The third layer is the actuation units, where the user can set the scaling and select the actuators. The fourth layer adjusts the control strategy. The user can influence the control parameters, including the objective function and constraints of the NEMPC controller.

While the user interacts with the interface, it provides real-time calculations. The user can also simulate the greenhouse operation over a period of time and analyze the results in terms of energy consumption, crop yield, and economic output.

\section{Results}
In this study, the optimization and simulation were carried out using CasADi\cite{Andersson2019}, a powerful tool for symbolic algorithmic differentiation, along with the IPOPT solver\cite{Wachter2006} for solving the nonlinear program. CasADi was used to model and discretize the Nonlinear Economic MPC problem, while IPOPT efficiently handled the optimization. These tools enabled the dynamic simulations of the greenhouse control system, ensuring computational efficiency and scalability.

\begin{figure*}
    \centering
    \includegraphics[width=\textwidth]{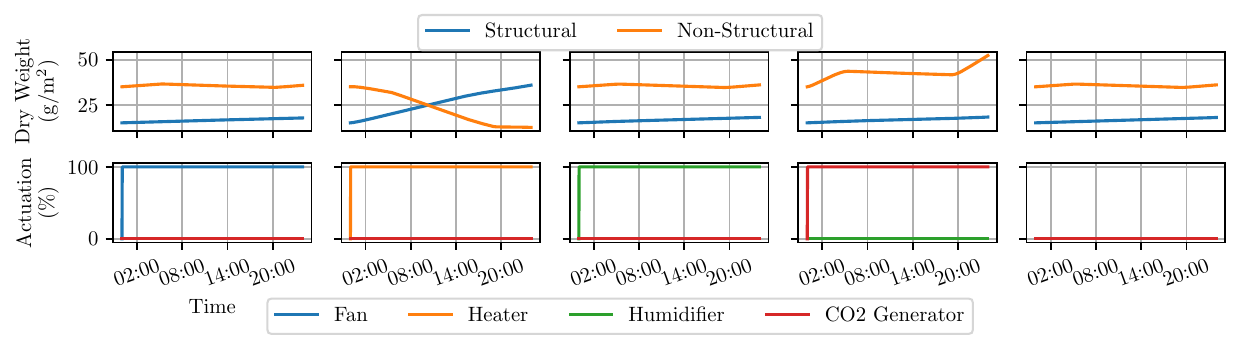}
    \caption{Responses of structural and non-structural dry weight on step changes from off to maximum in following actuations from the left to the right: ventilation, heating, humidification, and \coo\ enrichment, no step made.}\label{fig:steps}
\end{figure*}

First, we analyze the nonlinear behavior of the greenhouse under varying climate and actuation conditions. We observed how the actuators influence the growth of SDW and NSDW.\@ The simulations were run under a mild climate of Bratislava (Slovakia) on the day of 11\textsuperscript{th} October. Column 1 and 3 in Figure~\ref{fig:steps} show that the influence of ventilation and humidification is insignificant on growth as compared to no change in actuation in column 5.
Nevertheless, there is a positive effect on convection and the overall transfer of energy. Intense heating positively influences the conversion of NSDW into SDW, yet it does not influence the overall NSDW buildup, as shown in 2\textsuperscript{nd} column. 4\textsuperscript{th} column displays, that \coo\ enrichment has a significant impact on non-structural plant growth.

\begin{figure}\label{fig:control}
    \centering
    \includegraphics[width=\linewidth, trim=10 10 10 10]{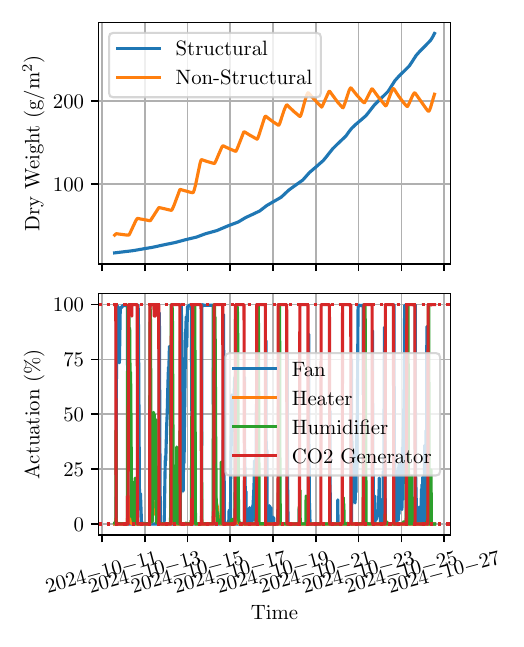}
    \caption{A simplified diagram of the heat, mass and \coo\ exchanges modeled within the framework. Image adapted from~\cite{rmward61_2019}.}
\end{figure}

The second set of simulations compares the performance of the proposed NEMPC algorithm with a no-control scenario. The NEMPC algorithm is configured with a prediction horizon and a control horizon of 1 hour. The simulations are run for a total of 10,801 steps, with a sampling time of 120 seconds. Results in Table~\ref{tab:comparison} demonstrate that the proposed NEMPC increased crop yield by 435\% and increased profit by more than 14\% in just 15 days of growth. As illustrated in Figure~\ref{fig:control}, which shows the lettuce growth and actuation patterns, the \coo\ generator is heavily relied upon, while the heater is used sparingly due to high electricity consumption and mild autumn weather.
The inclusion of the social cost of the carbon intensity into the cost function of the NEMPC leads to a 15\% reduction in \coo\ consumption, although it caused a 11\% decrease in plant growth and 3\% decrease in profit when compared to a scenario where the social cost of \coo\ was not minimized. This highlights a significant trade-off between economic output and the carbon intensity of the energy sources. While farmers may prioritize economic gains, the environmental impact of production deserves careful consideration.

\begin{table}
    \centering
    \caption{Performance comparison: NEMPC {vs. } no control.}\label{tab:comparison}
    \setlength{\tabcolsep}{4pt} 
    \begin{tabular}{lS[table-format=3.2]S[table-format=4.2]S[table-format=4.2]}
        \toprule
        \textbf{Parameter}  & {No control} & {NEMPC (\coo)} & {NEMPC (\$)} \\
        \midrule
        Lettuce profit      & 834.01       & 4459.57        & 4979.23      \\
        Energy (Fan)        & 0.00         & -0.34          & -0.36        \\
        Energy (Heater)     & 0.00         & -59.30         & -109.66      \\
        Energy (Humidifier) & 0.00         & -1.79          & -2.23        \\
        Energy (\coo\ Gen.) & 0.00         & -763.22        & -794.75      \\
        Energy (Solver)     & 0.00         & -0.001         & -0.001       \\
        \coo\ (Fan)         & 0.00         & -0.79          & -0.83        \\
        \coo\ (Heater)      & 0.00         & -158.51        & -378.51      \\
        \coo\ (Humidifier)  & 0.00         & -3.90          & -5.78        \\
        \coo\ (\coo\ Gen.)  & 0.00         & -2521.36       & -2705.12     \\
        \coo\ (Solver)      & 0.00         & -0.003         & -0.004       \\
        \midrule
        \textbf{Total}      & 834.01       & 950.36         & 981.99       \\
        \bottomrule
    \end{tabular}
\end{table}

To assess the educational impact of this work, we conducted a survey among students who interacted with the application. The survey measured the participants' prior knowledge and their learning outcomes in four key areas: mathematical modeling, optimal process control, economic process control, and MPC.\@

The results based on five responses show a varied range of initial expertise levels in both modeling and process control, with participants rating their skills from novice (1) to advanced (5). Despite this variation, the application demonstrated a significant educational benefit across all experience levels.

\paragraph{Low-Skilled Users}
Participants with minimal prior knowledge reported moderate improvements in understanding mathematical modeling, optimal control, and economic control, with a strong positive impact noted for MPC.\@ These results suggest that the framework is accessible to beginners and helps building foundational knowledge.

\paragraph{High-Skilled Users}
More experienced participants rated the application as highly beneficial in all four categories. They noted substantial improvements in their understanding of complex control techniques, particularly in mathematical modeling and MPC, validating the educational potential of the framework for users with advanced backgrounds.

\paragraph{General Feedback}
The participants' qualitative feedback further highlights the application's potential. Participants emphasized that with additional information, the tool could be used by a broader audience, including industrial farmers, to design and optimize greenhouse placement in practical settings. This points to the dual benefit of the application: it not only enhances student learning but could also have real-world applicability in sustainable greenhouse design.

These findings suggest that the proposed framework effectively supports educational objectives, promoting understanding across a spectrum of learners. It can also be adapted for a broader use beyond educational contexts, providing valuable insights into the design of sustainable greenhouse systems.

\section{Conclusion}
This study introduces a framework that combines the Nonlinear Economic Model Predictive Control with a detailed mathematical model to control the greenhouse climates, aiming to optimize lettuce growth. The provided framework utilizes real-time weather forecasts and carbon intensity data, to adjust the environmental conditions, improving crop yields, energy efficiency, and reducing \coo\ emissions.
The main contribution of this work is the educational tool helping students to understand not only the control theory but also to connect it with practical, real-world challenges. Through a user-friendly web-based platform, students can engage with advanced control theories while exploring the complexity of balancing economic goals and sustainability.
Student feedback has proven that the framework helps both beginning and advanced automatic control students to deepen their understanding of modeling, process control, and MPC techniques. Future work will focus on including other automatic control approaches to enrich the learning experience further.

\bibliographystyle{IEEEtran}
\bibliography{main}

\end{document}